\documentclass[12pt,preprint,flushrt]{aastex}
\begin{document}

\title{Evidence for an Intermediate Mass Black Hole in NGC 5408 X-1}
\author{Tod E. Strohmayer$^1$ \& Richard F. Mushotzky$^1$}
\altaffiltext{1}{Astrophysics Science Division, NASA's Goddard Space
Flight Center, Greenbelt, MD 20771 email: tod.strohmayer,
richard.mushotzky@nasa.gov} 

%\affil{Laboratory for High Energy Astrophysics, NASA's Goddard Space Flight
%Center, Greenbelt, MD 20771}
%\email{stroh@clarence.gsfc.nasa.gov}
%\authoraddr{Laboratory for High Energy Astrophysics, Mail Code 662, NASA/GSFC
%Greenbelt, MD 20771}

\begin{abstract}

We report the discovery with XMM-Newton of correlated spectral and
timing behavior in the ultraluminous X-ray source (ULX) NGC 5408
X-1. An $\approx 100$ ksec pointing with XMM/Newton obtained in
January, 2008 reveals a strong 10 mHz QPO in the $> 1$ keV flux, as
well as flat-topped, band limited noise breaking to a power law. The
energy spectrum is again dominated by two components, a $0.16$ keV
thermal disk and a power-law with an index of $\approx 2.5$.  These
new measurements, combined with results from our previous January 2006
pointing in which we first detected QPOs, show for the first time in a
ULX a pattern of spectral and temporal correlations strongly analogous
to that seen in Galactic black hole sources, but at much higher X-ray
luminosity and longer characteristic time-scales.  We find that the
QPO frequency is proportional to the inferred disk flux, while the QPO
and broad-band noise amplitude (root mean squared, rms) are inversely
proportional to the disk flux.
%The factor of two
%variation in the QPO frequency accompanied by a more modest change in
%the power-law energy spectral index suggests that a comparison with
%Galactic black holes approaching ``saturation'' of their QPO frequency
% - spectral index correlation can be made.  
Assuming that QPO frequency scales inversely with black hole mass at a
given power-law spectral index we derive mass estimates using the
observed QPO frequency - spectral index relations from five
stellar-mass black hole systems with dynamical mass constraints.  The
results from all sources are consistent with a mass range for NGC 5408
X-1 from 1000 - 9000 $M_{\odot}$.  We argue that these are
conservative limits, and a more likely range is from 2000 - 5000
$M_{\odot}$.  Moreover, the recent relation from Gierlinski et
al. that relates black hole mass to the strength of variability at
high frequencies (above the break in the power spectrum) is also
indicative of such a high mass for NGC 5408 X-1.  Importantly, none of
the above estimates appears consistent with a black hole mass less
than $\approx 1000$ $M_{\odot}$ for NGC 5408 X-1. We argue that these
new findings strongly support the conclusion that NGC 5408 X-1 harbors
an intermediate mass black hole.

\end{abstract}


\vfill\eject

%\doublespace

\section{Introduction}

The nature of the bright X-ray sources found in nearby galaxies, the
ultraluminous X-ray sources (ULXs), remains a major astrophysical
puzzle.  The fundamental conundrum is that some of these objects have
X-ray luminosities uncomfortably high to be stellar-mass black holes
(BH) without violating standard Eddington limit arguments. Three
different solutions have been proposed for the luminosity problem. 1)
The objects are intermediate-mass BHs (Colbert \& Mushotzky 1999). 2)
They are stellar-mass BHs with, in some cases, substantial beaming of
their X-ray radiation (King et al. 2001), or, 3) they are stellar-mass
BHs emitting above their Eddington limit (Begelman 2006).  It is
possible that some ULXs appear ultraluminous because of a combination
of all three factors (moderately higher mass, mild beaming and mild
super-Eddington emission). It may also be that these objects make up
an inhomogeneous population, comprised of both a sub-sample of
intermediate-mass BHs and moderately beamed stellar BHs (for recent
reviews see Fabbiano \& White 2006; Miller \& Colbert 2004).  Due to
their extragalactic nature the study of ULX counterparts at other
wavebands has been difficult, and has so far precluded the use of the
familiar methods of dynamical astronomy to weigh them. However, recent
work has resulted in mass measurements for some Local Group stellar
BHs, including IC 10 X-1 (Prestwich et al. 2007; Silverman \&
Filippenko 2008), and M33 X-7 (Orosz et al. 2007; Liu et al. 2008),
but these are not ULXs.

A substantial body of work now supports the idea that timing
properties can be used to constrain the masses of BHs. For example,
McHardy et al. (2006) have shown that when luminosity variations are
taken into account, both stellar-mass and supermassive BHs populate a
``variability plane'' linking their broad band variability
time-scales, luminosities and masses (see also K{\"o}rding et
al. 2007). Casella et al (2008) applied this relation in order to
estimate the masses of two ULXs, M82 X-1 and NGC 5408 X-1, concluding
that the masses of NGC 5408 X-1, and M82 X-1 lie in the range (in
solar units) $115 < M_{bh} < 1300$, and $95 < M_{bh} < 1260$,
respectively. However, in this case the scaling was not direct in that
Casella et al. (2008) had to estimate the relevant break time-scale
from the observed QPO frequencies in M82 X-1 and NGC 5408 X-1. They
did this by applying a scaling relation between the two quantities
derived only from observations of stellar-mass BHs.

Recently, Gierlinski, Nikolajuk \& Czerny (2008) have shown that the
strength of high frequency variability--parameterized as the root mean
squared (rms) amplitude integrated above the break in the power
spectrum--scales approximately linearly with BH mass.  They show that
a broad correlation exists when comparing stellar and supermassive
BHs, but the relation is not tight enough to predict the mass of a
stellar system by comparison, for example, with another stellar-mass
BH of known mass.

It is now firmly established that Galactic BHs accreting at high
rates--often classified as the Intermediate State (IS) or Steep
Power-Law state (SPL)--show strong correlations between their spectral
and temporal parameters.  In these states a significant fraction of
the X-ray luminosity is in a power-law component whose spectral index
correlates well with the frequency of a QPO (so-called Type C QPOs)
that is also commonly detected in such states (see Sobczak et
al. 2000; Vignarca et al. 2003; Kalemci et al. 2005; Remillard et
al. 2002; Casella et al. 2004). Other spectral parameters also
correlate with QPO frequency, such as the disk flux.  Recent work by
Shaposhnikov \& Titarchuk (2007; 2009; hereafter ST07 and ST09) has
demonstrated rather convincingly that the QPO frequency - spectral
index scaling for Type C QPOs can be used as an empirical BH mass
estimator. Previous efforts have been made to use this scaling
argument to constrain the masses of ULXs (Fiorito \& Titarchuk 2004;
Dewangan, Titarchuk \& Griffiths 2006; Strohmayer et al. 2007,
hereafter Paper 1), however, in these previous studies, there was no
direct evidence to indicate that the QPO properties seen in the ULX
sources actually correlates with the spectral parameters as in
stellar-mass systems.

In this paper we present the results of new timing and spectral
measurements that show for the first time that NGC 5408 X-1 behaves
very much like a Galactic stellar-mass BH system with the exception
that its characteristic X-ray time-scales are $\sim 100$ times longer,
and its luminosity is greater by a roughly similar factor.  We argue
that these new findings provide strong evidence that this system
contains a few thousand solar mass BH.

\section{New XMM-Newton Observations and Data Analysis}

Paper 1 summarizes results from XMM-Newton observations obtained in
2006 January in which X-ray QPOs were detected from NGC 5408 X-1 for
the first time (we sometimes also refer to this as Observation 1).  In
order to look for correlated timing and spectral behavior we sought
and obtained additional XMM-Newton observations in 2008 January. These
new observations began on January 13, 2008, and continued for $\approx
116$ ksec (Observation 2). We used the standard SAS version 8.0.0
tools to filter and extract images and event tables for both the pn
and MOS cameras.  We extracted events in an 18'' radius around the
source in both the pn and MOS cameras. Due to its higher count rate
and better time resolution we began our timing study using the pn
data.

\subsection{Power Spectral Timing Analysis}

There was some background flaring present during our observation which
broke up the exposure into several useful intervals.  We work with the
four longest intervals, labeled A through D in time order, and with
exposures of 15.9, 23.0, 20.0 and 13.6 ksec, respectively. Figure 1
shows the pn lightcurve in 78 s bins for the longest interval
(interval B).

We began our timing analysis by making average power spectra by
combining all good intervals.  Based on the energy dependent
variability behavior exhibited in observation 1 (see Paper 1) we made
power spectra in several energy bands; 0.2 - 8 keV, 1 - 8 keV, and 2 -
8 keV.  We found a prominent peak at 10 mHz in the $1 - 8$ keV power
spectrum. Figure 2 shows the average power spectrum binned at 0.64 mHz
in which we found the peak.  To assess the significance of the peak we
first rescaled the power spectrum. We did this by fitting a bending
power-law continuum to the spectrum, excluding the two highest bins in
the peak to avoid biasing the fit to higher values.  We then divided
the power spectrum by the best-fitting continuum model.  This
procedure is important when searching for QPOs against an intrinsic,
broad-band noise component as is present in NGC 5408 X-1.  We estimate
the significance as the chance probability of obtaining the highest
observed peak in the rescaled power spectrum using for the noise power
distribution the $\chi^2$ distribution with 176 degrees of freedom
(dof).  This is the expected distribution given the number of
independent frequency bins that were averaged (88). We compared the
distribution of powers in our observed power spectrum with this
expected distribution and confirmed they are consistent.  This then
gives a chance probability, per trial, of $3.7 \times 10^{-8}$. We
searched out to 0.5 Hz and this was the 2nd power spectrum searched,
giving a total of $390 \times 2 = 780$ independent trials, resulting
in a significance of $2.9 \times 10^{-5}$, a strong detection.

To further quantify the variability we fit the same continuum model
but now including a Lorentzian to account for the QPO. This model fits
well, with a best fitting $\chi^2 = 463$ for 459 degrees of freedom,
and is shown in Figure 2. Removing the QPO component from the fit
results in an increase in $\chi^2$ of 49. For the addition of three
additional QPO parameters, this increase in $\chi^2$ gives a chance
probability by the F-test of $5 \times 10^{-10}$, further supporting
the detection.  

The QPO has a centroid frequency of $10.24 \pm 0.005$ mHz, and a
coherence of $\nu_{cent} / \Delta\nu_{fwhm} \approx 20$.  The QPO is
strong, with an average amplitude (rms) in the $> 1$ keV band of $17.4
\pm 1.3 \%$.  We do not detect the QPO at energies below 1 keV, with a
$90\%$ upper limit on the amplitude (rms) of $4.5\%$, demonstrating
that the amplitude of the QPO is a strong function of energy.  The
band-limited noise is also quite strong, with an integrated amplitude
(rms) of $41 \pm 5 \%$ (1 - 100 mHz). Similarly to the QPO the
broad-band noise amplitude is also stronger at higher energies.  This
behavior is typical of Galactic BHs as well (Nowak et al. 1999;
Belloni et al. 2005).

We next examined the intervals separately. For the longest interval
(B) we again detect the 10 mHz QPO, but we also find evidence for a
second QPO at 13.4 mHz.  Figure 3 (upper curve) shows the power
spectrum from interval B in the $1 - 8$ keV energy band.  The 13.4 mHz
feature is significant at a bit better than the $3\sigma$ level, so we
consider it robust.  Interestingly, these two peaks have frequencies
consistent with a ratio of 4:3.  A pair of QPOs with a similar
frequency ratio were also detected in observation 1 (see paper
1). These similarities provide added confidence that the 13.4 mHz QPO
is significant.  Combining the two longest intervals (B and C) above 1
keV gives a power spectrum with a strong detection of the 10 mHz
feature, as well as two other candidate features (see Figure 3, lower
curve).  The higher frequency QPO at 13.4 mHz is still present, though
it is less prominent than in interval B alone, and a third feature is
suggested at 6 mHz.  We modeled this power spectrum with the same
bending power-law continuum and included up to three QPO components.
We find a statistically acceptable fit with $\chi^2 = 518.8$ (497
degrees of freedom). The 10 mHz QPO is absolutely required, as
removing it from the fit increases $\chi^2$ by 96, better than an
$8\sigma$ detection based on the F-test.  Removing the lowest and
highest frequency QPO components one at a time results in increases in
$\chi^2$ of 17.7 and 29.2, respectively.  These correspond to F-test
probabilities of $2.6 \times 10^{-4}$ and $6.9 \times 10^{-7}$,
respectively, for the additional components.  While this provides
rather strong evidence for the 13.4 mHz QPO, we regard the 6 mHz
feature as tentatively detected.

In summary, our new observations of NGC 5408 X-1 reveal a strong QPO
at 10 mHz, and very strong ``flat-topped'' band-limited noise breaking
to a power law, with the break close to the QPO.  This behavior is
qualitatively similar to results from our 2006 observations, but with
the QPO and break frequencies shifted down in frequency by about a
factor of two, and with stronger rms variability.

\subsection{Energy Spectral Analysis}

Previous spectral studies have shown that NGC 5408 X-1 has a cool
thermal disk component with $kT \approx 0.15$ keV and a power-law
extending to higher energies with a slope of $\approx 2.5$ (Kaaret et
al. 2003; Soria et al. 2004; Paper 1).  We obtained a pn spectrum by
extracting an 18'' region around the source.  Background was obtained
from a nearby circular region free of sources.  We began by fitting
the present spectrum with the same model components used to describe
our 2006 data (paper 1); a relativistic disk ({\it diskpn} in XSPEC),
a power-law, and a thermal plasma ({\it apec} in XSPEC).  These were
modified by successive photoelectric absorption components, one fixed
at the best Galactic value ($n_H = 5.7 \times 10^{20}$ cm$^{-2}$), the
other left free to account for possible local absorption.  This model
provides an acceptable fit, and from a qualitative standpoint the
spectral parameters are similar to those derived from our 2006
observations.  The disk temperature of $0.16 \pm 0.005$ keV is
consistent within the errors to that derived from the 2006 data.  The
power-law index, at $2.47 \pm 0.04$, is slightly smaller than the
previous measurement, but only different at about the 2$\sigma$ level.
We searched for line emission in the Fe band but did not detect any
significant features. However, the limits are not that constraining,
with an $90\%$ confidence upper limit on the equivalent width at 6.4
keV of $\approx 350$ eV. 

The thermal plasma model parameters and flux are consistent with the
2006 data, contributing a 0.3 - 10 keV flux of $\approx 1\times
10^{-13}$ erg cm$^{-2}$ s$^{-1}$ in each observation.  Examination of
{\it Chandra} images of NGC 5408 indicate the existence of extended
emission co-spatial with NGC 5408 X-1. This emission is unresolved in
the XMM observations of NGC 5408 X-1, and we suggest that this
spectral component is likely associated with star formation in NGC
5408 and not intrinsic to the ULX.  We will compare and constrast
the spectral results from the two epochs in more detail below.

\section{Comparison of 2006 and 2008 Observations: Timing - Spectral 
Correlations}

We can now compare the timing and spectral properties from the two
epochs, and determine whether or not NGC 5408 X-1 shows patterns of
behavior similar to that seen in Galactic BHs.  We begin with a more
detailed comparison of the energy spectrum and fluxes during the two
observations. For consistency we re-extracted the spectrum from the
2006 observations to include essentially all of the good exposure, and
in each epoch the spectrum represents the total accumulated over the
entire observation. We produced count rate spectra for each
observation and Figure 4 shows the difference spectrum as a function
of energy (Observation 1 - Observation 2).  One can see from the
figure that the first observation (2006) had a higher count rate, and
that most of the difference is in the $< 1$ keV band, which is
dominated by the disk flux.  This demonstrates that the 2006
observation had relatively more counts in the soft band ($< 1$ keV),
and less in the power-law component extending to higher energy.  Table
2 provides a detailed comparison of the spectral parameters derived
from the two epochs. The unabsorbed flux (0.3 - 10 keV) in 2008 was
$3.1 \times 10^{-12}$, about $12 \%$ less than the 2006 observation,
and the power-law component represents a greater fraction of the total
flux than in the 2006 data. The two most important conclusions are
that the disk flux was about $20 \%$ higher in Observation 1, and that
the power-law component contributes a greater fraction of the total
flux in Observation 2.

We next compared power spectra accumulated during each
observation. Figure 5 shows power spectra characteristic of each
epoch.  The upper curve is data from our 2008 observations, and the
lower curve is from 2006 (see also Paper 1, Figure 3).  These power
spectra were not accumulated over the exact same energy bands.
Because the variability has a significant energy dependence, and
because a primary goal is to compare the characteristic time-scales in
each epoch, we choose to compare the spectra in the energy bands where
the signal to noise ratio of the respective QPOs is largest.  For the
2006 data this corresponds to 0.2 - 8 keV, whereas for the 2008 data
we use, of the bands searched, the 1 - 8 keV band.  Table 1 compares
relevant timing properties for the two observations.  Timing
properties were derived from the model fits shown in Figure 5 using
the bending power-law continuum and Lorentzian components for the
QPOs.  We note that the bend frequency in the $0.2 - 8$ keV 2006 data
we used to compare the 2006 and 2008 QPO parameters is not well
constrained. A more representative value is the bend frequency of
$\approx 25$ mHz derived from the $> 2$ keV 2006 data (Paper 1).  In
general, the measurement of the QPO frequency in both epochs is much
more precise than the bend frequency, which is why we emphasize it in
our comparisons of the two epochs.

Closer examination of Figure 5 reveals several important conclusions
concerning the variability in NGC 5408 X-1.  We see that the strongest
QPO has shifted from 20 to 10 mHz, over the same time that the disk
flux dropped by $20 \%$.  Additionally, the photon power-law index
decreased modestly as the QPO frequency dropped. These behaviors are
entirely consistent with what has been observed in Galactic BHs with
so-called Type C QPOs (see, for example, Sobczak et al. 2000; Vignarca
et al. 2003; Kalemci et al. 2005).

Another strikingly evident feature in Figure 5 is the larger amplitude
of the variability (broad-band and QPO) in the 2008 observation (top).
This quantity is proportional to the integral of the power spectrum
above the poisson level (here a value of 2), which is clearly larger
in the 2008 observations.  The rms amplitude also has a scaling with
the square root of the count rate, but this was smaller in the 2008
observations, so this effect simply makes the apparent difference
larger.  We also compared the rms amplitudes {\it in the same energy
band}, and the conclusion is unchanged. Again, the higher rms
variability at the epoch of lower disk flux is strikingly similar to
what is observed in Galactic systems.  Here the primary conclusion is
that the variability is mostly carried by the power-law component, and
not the disk flux.  So, when the power-law becomes more dominant, then
the rms amplitude increases.  Finally, we note that the modest drop in
the power-law index with the decrease in QPO frequency is also
consistent with the correlations seen in Galactic systems.  In this
case the relatively small change in the index accompanying a larger
change in QPO frequency suggests that the power-law index in NGC 5408
X-1 may be near ``saturation'' of the scaling (see for example, Figure
5 in Vignarca et al. 2003; ST09), that is, both the QPO frequency and
index are near their upper ranges in the correlation.  This conclusion
is also supported by the behavior of the rms amplitude. The strong
drop in the rms amplitude at relatively constant power-law index is
consistent with the QPO frequency being at the higher end of the
observed correlations (see, for example, the behavior of XTE J1550-564
illustrated in Figure 5a from Sobczak et al. 2000).

\section{Discussion and Implications}

Our new observations of NGC 5408 X-1 reveal correlated variations in
its timing and spectral properties very much like a stellar-mass black
hole.  Indeed, the evidence is now quite compelling that the strong
low frequency QPO detected on several occasions from NGC 5408 X-1 is a
Type-C QPO analogous to those seen in Galactic systems.  It varies in
frequency and amplitude with changes in flux and spectrum in a manner
entirely consistent with the behavior in Galactic BHs such as XTE
J1550-564 and GRS 1915+105. Whereas these stellar-mass systems have
characteristic QPO frequencies of a few Hz, NGC 5408 X-1 shows QPO
frequencies lower by a factor $\sim 100$ while simultaneously
radiating an X-ray luminosity larger by a roughly similar factor.

Given greater confidence in the identification of and scaling
properties of the strong QPO seen in NGC 5408 X-1 we can use it to
derive a mass estimate.  We essentially follow the scheme outlined by
ST09.  We use the QPO frequency - power-law index correlations
measured for five systems; GRS 1915+105 (Vignarca et al. 2003), XTE
J1550-564, XTE J1859+226, H 1743-322 and GX 339-4 (ST09).  We use
these systems as our primary sources for mass estimates because they
have both measured power-law indices that overlap the range of
observed indices in NGC 5408 X-1, and reasonable BH mass estimates
from dynamical measurements.  

%Below we discuss what we think are less
%certain scalings derived from the other Galactic BH systems with
%measured correlations, but whose observed spectral indices are less
%than the range observed from NGC 5408 X-1.

Since we do not yet have a correlation from NGC 5408 X-1 with many
points to scale from we carry out a simplified procedure. For our
reference Galactic systems we find the range of QPO frequencies with
power-law indices between 2.4 and 2.6.  We choose this index range as
representative and conservative.  It is 2$\sigma$ below the lower best
fit value (from 2008) and 2$\sigma$ above the higher best-fit value
(from 2006) for NGC 5408 X-1. In the case that a range of QPO
frequency exists at either end of our index range, we take the largest
range of observed frequencies.  To be specific, we take the lowest
frequency measured at an index of 2.4, and the highest frequency found
at an index of 2.6. We then find the multiplicative scale factor, $f$,
required to bring the lower and upper QPO values measured for NGC 5408
X-1 (10 and 20 mHz) into agreement with the measured frequency range
for our reference systems. In the case that a constant scale factor
cannot match both the lower and higher frequencies of the observed
range, we find the scale factor that aligns the centers of the two
ranges (we call this estimate $f_{best}$). The derived mass estimate
for NGC 5408 X-1 is then the mass of the reference stellar system
times the scale factor.  In one case, H 1743-322, frequency
measurements exist only near the low end of our target index range
(2.4).  In this case we derive the scale factor by simply scaling to
the lower QPO frequency measured in NGC 5408 X-1.  The derived scale
factors and mass estimates based on comparison with these five
reference systems are summarized in Table 3.

The quoted uncertainty on the best mass estimate in Table 3 reflects
the uncertainty in the masses of the reference systems (where
available), that is, assuming a ``correct'' scale factor.  The QPO
frequency and power-law index measurements for NGC 5408 X-1 described
here are rather precise, so that if all the assumptions made with
regard to the mass scaling arguments above are accurate, then the
amount of statistical error in the estimates is modest.  This leaves
systematic uncertainties, that is, how accurate is the derived scale
factor?  Now, some Galactic BHs do show variations in their QPO
frequency -- spectral index correlations.  These are primarily
associated with apparent changes in the value of the spectral index at
which the correlation saturates (see, for example, the behavior of XTE
J1550-564 in Figure 7 from ST09), or simply that the correlation is
not exact, and that at a given index value, there can be a range of
measured QPO frequencies.  We took into account such variations in
defining the range of QPO frequencies to scale to for each reference
system.  However, the existence of a range of measured frequencies at
a given spectral index suggests a conservative way to bound the scale
factor sytematic error, by defining a minimum and maximum scale factor
given the observed QPO frequency range. We do this by scaling the
lowest observed QPO frequency in our reference systems with the
highest observed QPO frequency in NGC 5408 X-1, and vice versa. For
example, for XTE J1550-564, which has observed QPO frequencies from
2.5 - 6.5 Hz, the minimum and maximum scale factors would be (2.5 Hz)
/ (0.02 Hz) = 125, and (6.5 Hz) / (0.01 Hz) = 650.  Based on these
limits on the scale factor we also derive minimum and maximum mass
estimates by multiplying the maximum and minimum scale factors by the
$\pm 1\sigma$ mass limits for each reference system.  Both the scale
factor ranges and minimum and maximum mass estimates are also given in
Table 3.  We emphasize that we think these represent rather
conservative limits.

While the derived mass ranges in Table 3 are rather large--and we
emphasize that this is not a precision mass measurement
technique--there is substantial overlap among the estimates from all
the different sources. Perhaps more interesting is that all the
scaling estimates suggest a BH with a mass comfortably greater than
$1000 M_{\odot}$, that is, much greater than the current known mass
range for stellar BHs.  We note that the candidate BH system 4U
1630-47--which is not a dynamically confirmed BH--also has QPO
frequency measurements that overlap our target range of spectral index
(see Figure 9 in Vignarca et al. 2003). In this source the frequency
range is $\approx 4 - 8$ Hz, suggesting a best scale factor of 400,
which falls in the range derived for our other reference systems.
While it's mass is not known, a typical value of $\sim 10$ $M_{\odot}$
would also suggest a mass for NGC 5408 X-1 in the range of several
thousand $M_{\odot}$, consistent with the other sources.

\subsection{Other Concerns and Caveats}

One concern with regard to QPO frequency scaling arguments has been
which QPO frequency to scale to. For example, both XTE J1550-564 and
GRS 1915+105 can sometimes show QPO frequencies below 1 Hz,
approaching 0.1 Hz in the case of XTE J1550-564. However, our
observations of QPOs from NGC 5408 X-1 at different epochs, and with
different frequencies and rms amplitudes goes along way towards
alleviating this concern. In fact, the behavior of the QPO rms
amplitude in NGC 5408 X-1 is completely reversed to what one would
expect for a scaling with the lowest QPO frequencies observed in XTE
J1550-564 and GRS 1915+105. Results from XTE J1550-564 show that at
low QPO frequency, the rms increases with an increase in QPO frequency
(see Figure 5a in Sobczak et al. 2000), but this is exactly opposite
to the behavior seen in NGC 5408 X-1. The behavior of the QPOs in both
XTE J1550-564 and GRS 1915+105 when the power-law index and frequency
are both high is a much better match to the behavior seen in NGC 5408
X-1.

Another concern has been that while QPO frequency scalings may give
one result, a simple scaling of luminosities gives another, and that
the two may not be in agreement.  Here one must try to be consistent
and compare luminosities under the same conditions.  In the case of
our scaling arguments this would be to match luminosities at the
appropriate power-law index and scaled QPO frequency.  An important
issue here is that in many cases the distances to BHs in the Galaxy
are relatively less well known than the distances to nearby galaxies
hosting ULXs. Nevertheless, we can attempt to compare the luminosity
of NGC 5408 X-1 to some of our reference BHs.  For XTE 1550-564, ST09
report X-ray spectral and QPO frequency measurements.  For power-law
index and QPO frequency appropriate to our scaling the observed X-ray
flux from XTE J1550-564 (2 - 20 keV) was in the range 3 - 5 $\times
10^{-8}$ erg cm$^{-2}$ s$^{-1}$.  The distance to this source is not
very well constrained, with estimates ranging from 2.5 - 6 kpc.
Taking a flux of $4\times 10^{-8}$ erg cm$^{-2}$ s$^{-1}$ as
representative, we have a luminosity of $5 \times 10^{37} (d/{\rm 3
kpc})^2$. This compares with a representative luminosity from our 2006
observation of about $1.1 \times 10^{40}$ erg cm$^{-2}$ s$^{-1}$ (0.2
- 10 keV; assuming a distance of 4.8 Mpc). The luminosity ratio is
then in the range of 317 - 55, for the distance range of 2.5 - 6 kpc
for XTE J1550-564. Simply scaling up the $\pm 1\sigma$ mass range for
XTE J1550-564 by these limits gives a mass estimate for NGC 5408 X-1
of $462 - 3360$ $M_{\odot}$.  Interestingly, ST09 use an observed
correlation in the power-law index and bulk motion comptonization
model normalization (model {\it bmc} in XSPEC) to estimate distances
as well as BH masses by scaling to a reference system.  Interestingly,
they favor a distance closer to 3 kpc than 6 kpc for XTE J1550-564,
which would favor the high end of the derived mass range.  For GRS
1915+105, again the distance is rather uncertain, likely being in the
range from $6 - 12$ kpc (Dhawan et al. 2007).  For the QPO frequency
range above a representative flux range is about 2 - 5 $\times
10^{-8}$ erg cm$^{-2}$ s$^{-1}$ (Muno et al. 1999), giving a range of
luminosity ratio of $73 - 18.2$.  Similar scaling as for XTE J1550-564
would imply a mass range from $182 - 1314$ $M_{\odot}$. While this
range appears systematically smaller than the mass range inferred from
the QPO scaling, it still overlaps with the minimum mass estimate for
GRS 1915+105.

\subsection{Amplitude of High Frequency Variability}

Recently, Gierlinski et al. (2008) have argued that the amplitude of
X-ray variability at high frequencies can be used as an estimator of
BH mass.  Their hypothesis is that there is a ``universal'' power
spectral shape for BHs at high frequency. Here, high frequency means
above the break in the power spectrum.  This universal form is roughly
a power law with index of 1.5 - 2.0.  They explore the notion that
this part of the power spectrum is nearly constant for a given source,
but scales with BH mass.  They find that there is a roughly linear
correlation extending from the stellar-mass BHs to AGN (using type 1
Seyferts).  The relation is not exact, and the scatter is such to make
it difficult to estimate the mass of a stellar BH by scaling from
another stellar-mass system, however, for order of magnitude estimates
the method seems reasonably robust.  We used our power spectrum
continuum models (above the bend or break in the spectrum) to estimate
the parameter $C_{M}$, which is simply the normalization of the high
frequency power-law component at 1 Hz.  We integrated our best fitting
continuum models from the bend frequency to 0.5 Hz in order to
estimate $C_{M}$ (see Eqn. 2 in Gierlinski et al. 2008). We find $\pm
1\sigma$ ranges of $-3.23 < \log C_{M} < -3.13$, and $-3.66 < \log
C_{M} < -3.55$ for the 2008 and 2006 observations, respectively.
These values correspond to mass ranges of 1686 - 2612, and 4435 - 5714
$M_{\odot}$, respectively, for the best fitting correlation derived by
Gierlinski et al., and 3737 - 4704, and 9828 - 12661 $M_{\odot}$ using
their soft-state relation derived from Cyg X-1.  While the possible
mass range from this method is large, the $C_{M}$ values for NGC 5408
X-1 fall almost midway between the stellar-mass systems and NGC 4395,
a low mass AGN (see Figure 7 in Gierlinski et al. 2008), and
comfortably within the mass range consistent with intermediate mass
BHs.

\section{Summary and Conclusions}

Our new XMM-Newton observations have revealed for the first time that
NGC 5408 X-1 exhibits correlated X-ray timing and spectral properties
quite analogous to those exhibited by Galactic stellar-mass BHs in the
``very high'' or ``steep power-law'' state.  Its longer observed
time-scales (QPO and power spectral break frequencies) at higher
luminosity compared to Galactic stellar-mass BHs can be understood if
the mass of NGC 5408 X-1 is a few thousand solar masses.  We have
arrived at this conclusion by several independent arguments. Given our
present understanding of the timing properties of BHs at all mass
scales, it seems to us hard to escape the conclusion that NGC 5408 X-1
is an intermediate mass BH.

We have again found evidence for a pair of sharp, closely spaced QPOs
in NGC 5408 X-1 with a frequency ratio consistent with 4:3.  We thus
conclude that this is an intrinsic feature of the system and not some
artifact or coincidence associated with having only a single
observation of the source.  As we noted in Paper 1, it is possible
that this results from detection of both a Type C QPO (the stronger
feature), and a Type B QPO (Casella et al. 2004).  If the source
transitions from one to another, we do not have sufficient signal to
noise ratio data to ``watch'' such transitions occur. Rather, we
simply see average detections of each QPO.

We have argued that the observed variations in QPO frequency,
accompanied by spectral changes consistent with behavior seen in
Galactic systems, allows the mass of NGC 5408 X-1 to be estimated by
scaling the observed QPO frequencies to match those observed in
stellar-mass systems of known mass. Essentially the same method has
now been used on a good number of Galactic systems and seems
empirically robust (ST09).  These arguments give mass estimates for
NGC 5408 X-1 in a broad range from $1 - 9 \times 10^{3}$ $M_{\odot}$,
however, we find that none of the QPO scaling estimates are consistent
with masses below $\approx 1000$ $M_{\odot}$. Independent mass
constraints based on the amplitude of high frequency variability also
appear consistent with this range. Additional observations would help
to map out the timing - spectral correlations for NGC 5408 X-1 more
clearly, and thus allow more rigorous estimates.  As noted in Paper 1,
the measured disk temperature in NGC 5408 X-1 is also consistent with
a few thousand $M_{\odot}$ BH if the theoretical accretion disk
scaling of $kT_{disk} \propto M^{-1/4}$ holds. Importantly, none of
these new estimates appears consistent with a mass as low as even a
few hundred solar masses.  We think these new findings provide strong
evidence that NGC 5408 X-1 harbors an intermediate mass BH.

\acknowledgements

We thank the anonymous referee for a careful review of the manuscript
that helped us to improve the paper. 

\vfill\eject

\vfill\eject

\newpage

\begin{figure}
\begin{center}
 \includegraphics[width=6.5in, height=6in]{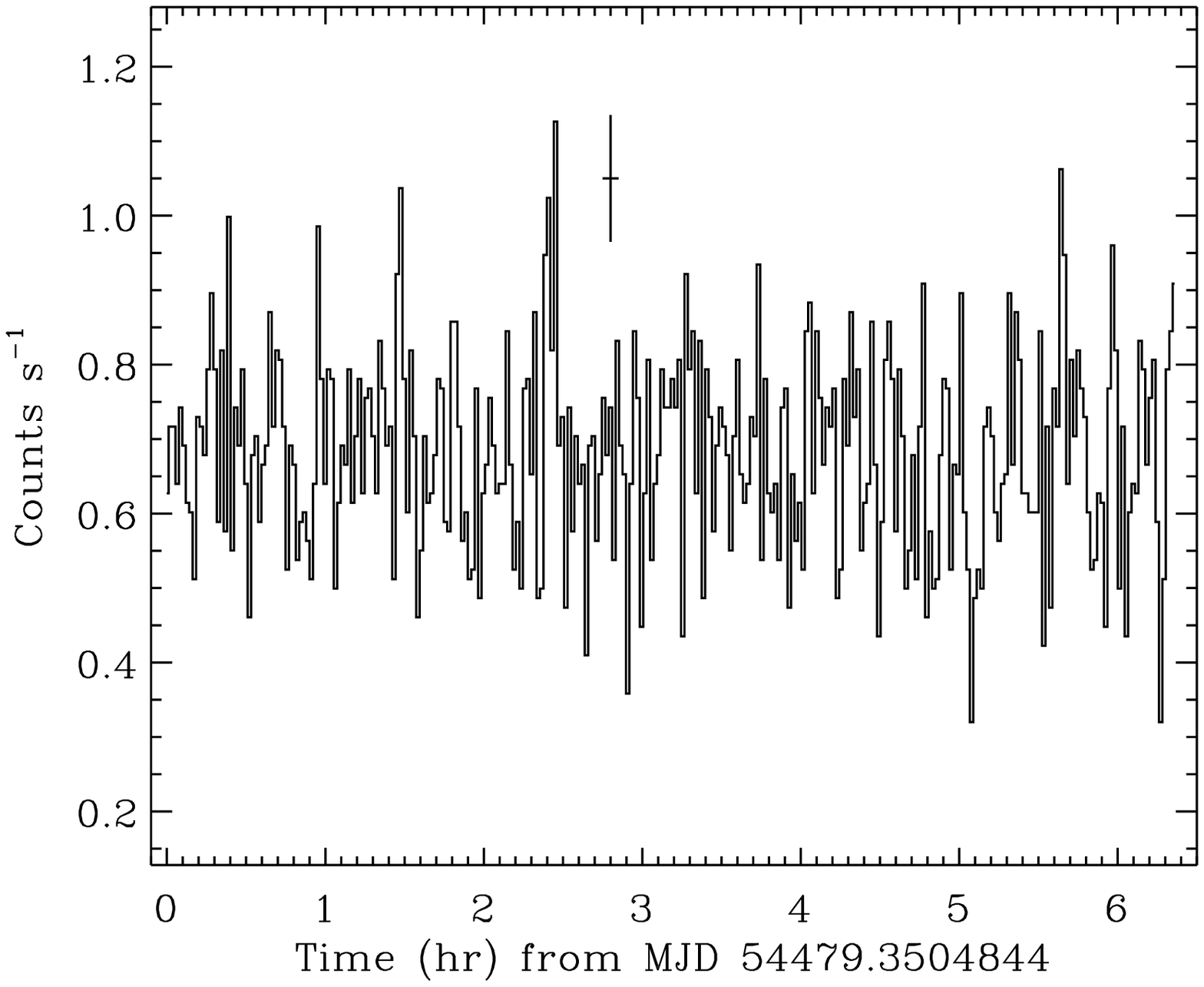}
\end{center}
Figure 1: Lightcurve of NGC 5408 X-1 (0.2 - 15 keV band) from
XMM-Newton EPIC/pn observations showing the longest contiguous time
interval used in our power spectral analysis (Interval B). The bin
size is 78.125 seconds. A characteristic error bar is also shown.
\end{figure}

\pagebreak

\begin{figure}
\begin{center}
 \includegraphics[width=6.5in, height=6in]{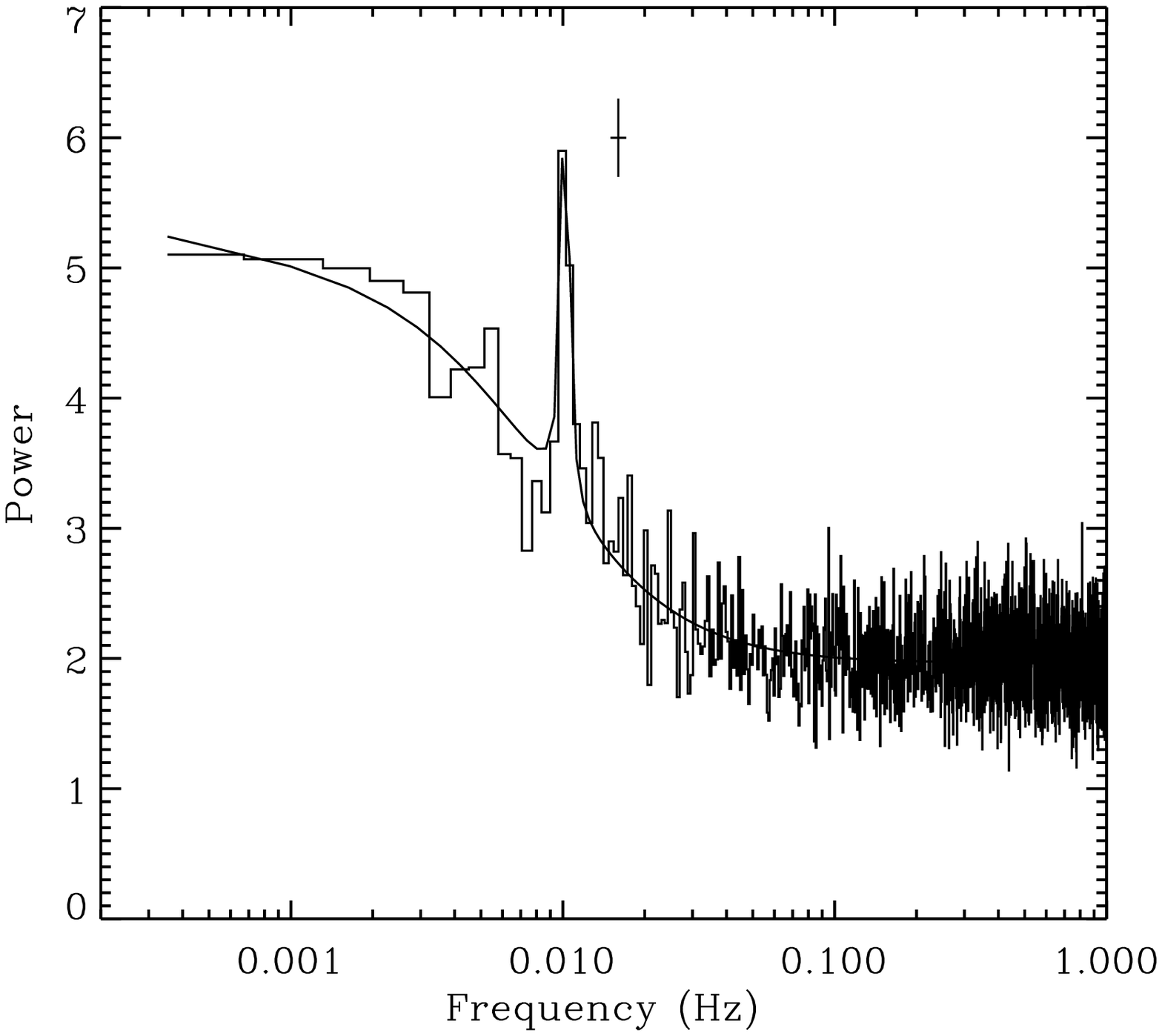}
\end{center}
Figure 2: Average power spectrum of NGC 5408 X-1 from the $> 1$ keV
EPIC/pn data (histogram) and the best fitting model (solid). The
frequency resolution is 0.64 mHz, and each bin is an average of 44
independent power spectral measurements.  The effective exposure is
$\approx 70$ ksec. A characteristic error bar is also shown. See the
text for a detailed discussion of the model, and Table 1 for model
parameters.
\end{figure}

\pagebreak

\begin{figure}
\begin{center}
 \includegraphics[width=6.5in, height=6in]{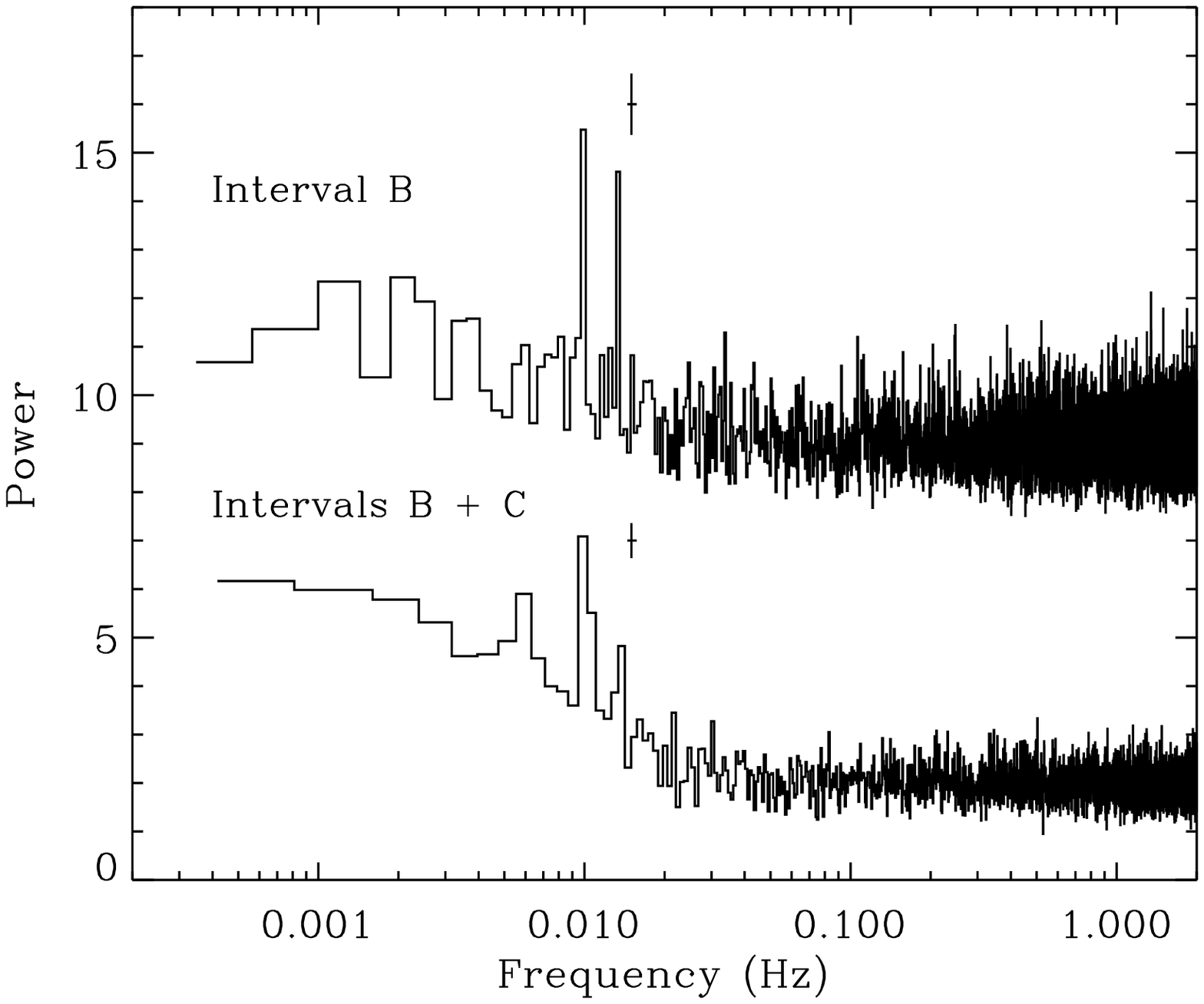}
\end{center}
Figure 3: Average power spectrum of NGC 5408 X-1 from $> 1$ keV
EPIC/pn data. The upper curve shows data from interval B with a
frequency resolution of 0.434 mHz. Each bin is an average of 10
independent measurements. The bottom curve is an average of the two
longest intervals (B and C) at a frequency resolution of 0.78
mHz. Each bin is an average of 32 independent measurements. See the
text for additional discussion.
\end{figure}

\pagebreak

\begin{figure}
\begin{center}
 \includegraphics[width=6.5in, height=6in]{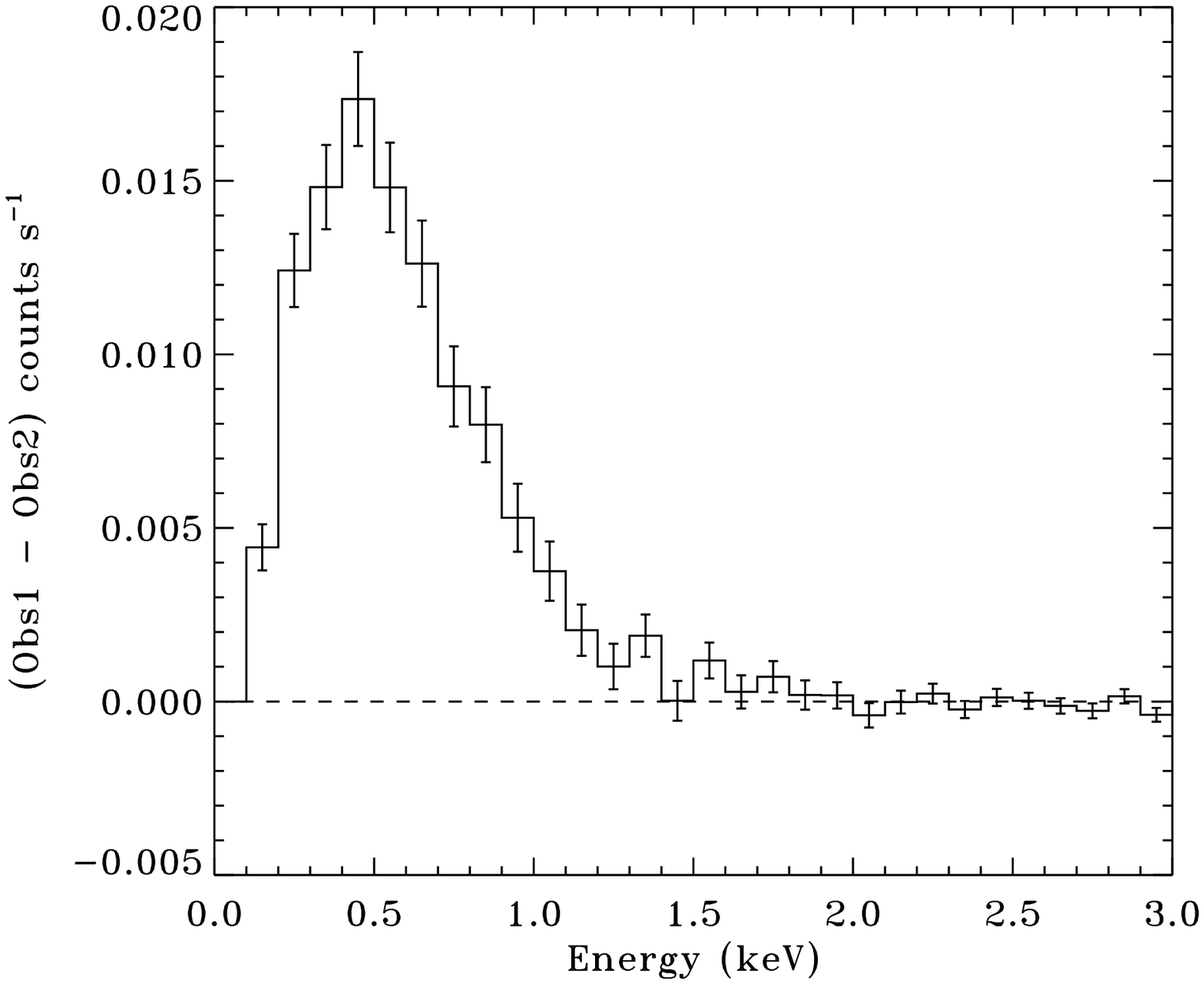}
\end{center}
Figure 4: Difference of the count rate spectra between Observation 1
and Observation 2 (i.e. Obs1 - Obs 2).  This shows that observation 1
was brighter, and that most of the difference was associated with $<
1$ keV photons, that is, the disk component.  
\end{figure}

\pagebreak

\begin{figure}
\begin{center}
 \includegraphics[width=6.5in, height=6in]{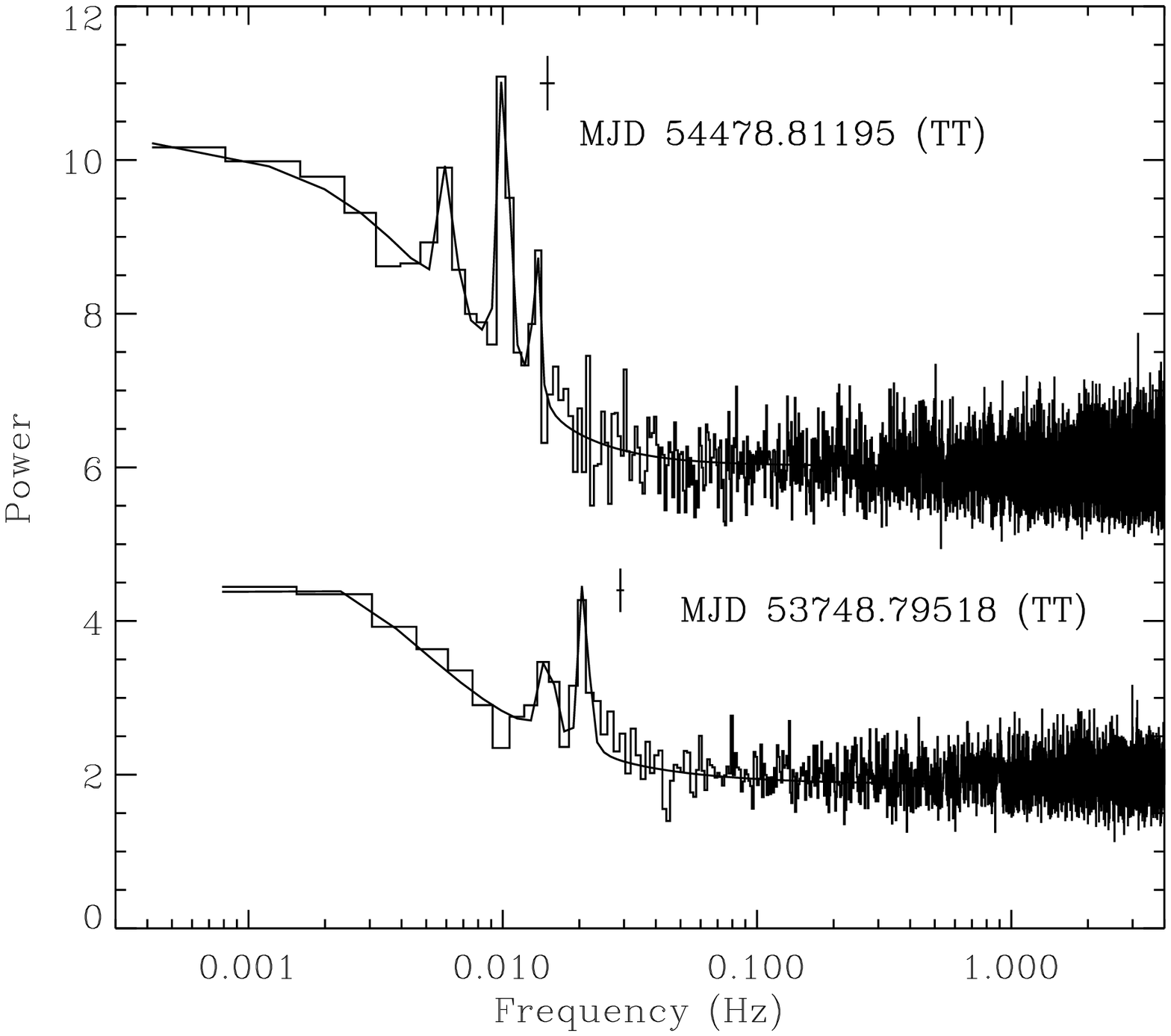}
\end{center}
Figure 5: Average power spectra and best fitting models characteristic
of both observations of NGC 5408 X-1 from EPIC/pn data.  The upper
curve is from the 2008 data (intervals B + C), while the bottom curve
is from the 2006 data. See the text for additional details and Table 1
for a summary of the model parameters.

\end{figure}
\pagebreak

%\typesize{\small}

\begin{deluxetable}{cccc}
\tablecaption{Results of Power Spectral Modeling for NGC 5408 
X-1\tablenotemark{1}}
\tablehead{\colhead{Parameters} & \colhead{2008, All intervals (Fig. 2)} & 
\colhead{2008, Interval B+C (Fig. 5)} & \colhead{2006 (Fig. 5)} } 

\startdata
 
A\tablenotemark{a} & $1.88 \pm 0.2$ & $3.4 \pm 0.6$ & $55.7 \pm 5$ \\[6pt]

$\alpha_L$\tablenotemark{b} & $0.07 \pm 0.08$ & $0.029 \pm 0.1$ & 
$-0.42 \pm 0.5$\\[6pt]

$\alpha_H$\tablenotemark{c} & $1.83 \pm 0.3$ & $1.7 \pm 0.4$ & $1.2\pm 0.4$ \\

$\nu_{{\rm bend}}$ (mHz)\tablenotemark{d} & $10.0 \pm 3$ & $6.3 \pm 2$ & 
$2.6 \pm 2$ \\

$r_{BB}$ ($\%$)\tablenotemark{e} & $41 \pm 5$ & $45 \pm 4$ & $23 \pm 4$ \\

$\nu_1$ (mHz)\tablenotemark{f} & $10.02 \pm 0.005$ & $10.01 \pm 0.002$ &
$20.6 \pm 0.03$ \\

$r_1$ $(\%)$\tablenotemark{g} & $17.4\pm 1.3$ & $15.0\pm 1.2$ & $8.4\pm 1.0$ \\

$\nu_2$ (mHz) & NA & $13.5 \pm 0.1$ & $15.1 \pm 0.2$ \\

$r_2$ $(\%)$ & NA & $10.0\pm 2$ & $5.4\pm 1$ \\

$\nu_3$ (mHz) & NA & $6.0 \pm 0.2$ & NA \\

$r_3$ $(\%)$ & NA & $8.8\pm 2.5$ & NA \\

$\chi^2$ (dof) & 463 (459) & 518.8 (497) & 152.2 (180) \\

\enddata \tablenotetext{1}{Summary of best fit power spectral models
for NGC 5408 X-1.  The results from fits to three different power
spectra are shown in columns 2-4.  Columns 2-4 show results using pn
data, but for different time intervals. The particular time interval
used is given with a reference in the heading to the figure the power
spectrum appears in.  These fits used up to three Lorentzian
components, numbered 1-3 in order of increasing rms amplitude.}
\tablenotetext{a}{Normalization of the bending power-law component.}
\tablenotetext{b}{Power law index below the bend frequency.}
\tablenotetext{c}{Power law index above the bend frequency.}
\tablenotetext{d}{Bend frequency, in mHz.}
\tablenotetext{e}{rms amplitude of the broad-band continuum}
\tablenotetext{f}{Centroid frequency of the strongest QPO}  
\tablenotetext{g}{rms amplitude of the strongest QPO}
\tablenotetext{h}{Centroid frequency of the next strongest QPO}  
\tablenotetext{i}{rms amplitude of the next strongest QPO}
\tablenotetext{j}{Centroid frequency of the weakest QPO}  
\tablenotetext{k}{rms amplitude of the weakest QPO}
\tablenotetext{l}{Minimum $\chi^2$ of the fit}

\end{deluxetable}

\pagebreak
%\insert{Table_ngc5408.tex}
\begin{deluxetable}{ccc}
%\tabletypesize{\scriptsize}
\tablecaption{Spectral Fits to XMM-Newton pn Spectra\tablenotemark{*}
\label{tbl-2}}
%\tablewidth{0pt}
\tablehead{
\colhead{Spectral parameters} &  \colhead{2006} & \colhead{2008}
}
\startdata
\multicolumn{3}{c}{Model: {\tt tbabs}*({\tt diskpn} + {\tt apec} 
+ {\tt pow})} \\
n$_H$\tablenotemark{a} & $7.0 \pm 0.5$ & $6.5 \pm 0.6$ \\
T$_{max}$\tablenotemark{b} & $0.155 \pm 0.004$ & $0.160 \pm 0.005$ \\
kT\tablenotemark{c} & $0.87 \pm 0.05$ & $0.85 \pm 0.04$ \\
$\Gamma\tablenotemark{d}$ & $2.58 \pm 0.04$ & $2.47 \pm 0.04$ \\
$\chi^2$/dof & 814/673  & 776/673 \\
F$_{disk}$ (0.3-10, keV)\tablenotemark{e} & $1.5\times 10^{-12}$ & $1.1 
\times 10^{-12}$ \\
F$_{pow}$ (0.3-10, keV)\tablenotemark{f} & $2.1 \times 10^{-12}$ & $ 1.9 
\times 10^{-12}$ \\
F$_X$ (0.3-10\,keV)\tablenotemark{g} & $3.7\times10^{-12}$ & $3.1 
\times 10^{-12}$ \\
\enddata
\tablenotetext{a}{Hydrogen column density in units of 10$^{20}$\,cm$^{-2}$, 
not including the Galactic contribution of n$_{H Gal} = 
5.73\times10^{20}$\,cm$^{-2}$.}
\tablenotetext{b}{Disk temperature in keV from the XSPEC disk model 
{\tt diskpn}. The inner disk radius was fixed at 6\,GM/c$^2$.}
\tablenotetext{c}{Plasma temperature in keV from the XSPEC model {\tt apec}.  
The abundances were fixed to the solar values.}
\tablenotetext{d}{Power-law spectral index.}
\tablenotetext{e}{Unabsorbed flux from the {\tt dispn} component, in units of 
erg\,cm$^{-2}$\,s$^{-1}$.}
\tablenotetext{f}{Unabsorbed flux from the power-law component, in units of 
erg\,cm$^{-2}$\,s$^{-1}$.}
\tablenotetext{g}{Unabsorbed total flux in units of erg\,cm$^{-2}$\,s$^{-1}$.}
\tablenotetext{*}{All errors are quoted at the 90\% confidence level.}
\end{deluxetable}

\pagebreak

\begin{deluxetable}{cccccc}
\tablewidth{1.1\columnwidth}
%\rotate
%\tabletypesize{\scriptsize}
\tablecaption{Mass Estimates for NGC 5408 X-1 from QPO Scaling\tablenotemark{*}
\label{tbl-3}}
\tablehead{ \colhead{Source} & \colhead{$M_{dyn}$\tablenotemark{a}} &
\colhead{$\nu_{qpo}$ range\tablenotemark{b}} & \colhead{$f_{min}$,
$f_{best}$, $f_{max}$\tablenotemark{c}} & \colhead{$M_{min}$,
$M_{best}$, $M_{max}$\tablenotemark{d}} &
\colhead{Refs.\tablenotemark{e}} \\ & $M_{\odot}$ & Hz & & $M_{\odot}$
& } \startdata XTE J1550--564 & $9.5 \pm 1.1$ & 2.5 - 6.5 & 125, 300,
650 & 1050, $2850\pm 330$, 6890 & 1, 2, 3 \\ GRS 1915+105 & $14 \pm 4$
& 2.1 - 3.5 & 105, 185, 350 & 1050, $2590 \pm 740$, 6300 & 2, 3, 4 \\
H 1743--322 & $\sim 11$ & 3 - 5 & 150, 267, 500 & 1650, 2937, 5500 &
3, 5 \\ XTE J1859+226 & 7.6 - 12.0 & 7.5 & 375, NA, 750 & 2850, NA,
9000 & 3, 6, 7 \\ GX 339--4 & $> 6$ & 5.9 - 7.8 & 295, 457, 780 &
1770, 2742, 4680 & 3, 8, 9 \\ 4U 1630--47\tablenotemark{f} & NA & 4 -
8 & 200, 400, 800 & NA & 2 \\ \enddata \tablenotetext{a}{Mass
estimates for dynamically confirmed BHs.}  \tablenotetext{b}{Range of
QPO frequencies observed when the spectral index ranges from 2.4 to
2.6.}  \tablenotetext{c}{Scale factors derived from the observed QPO
range. The ``best'' scale factor is derived by aligning to the center
of the range. The minimum and maximum values are obtained by scaling
the lowest observed frequency in the reference system with the highest
observed frequency in NGC 5408 X-1, and vice versa. See \S 4 for
discussion.}  \tablenotetext{d}{Estimated masses determined by scaling
up the observed masses of the reference systems by the derived scale
factors. The ``best'' mass estimate is obtained using the best scale
factor, with uncertainties set by the uncertainties in the reference
source masses. The minimum and maximum mass estimates are derived
using the minimum and maximum scale factors and the $\pm 1\sigma$ mass
limits. For example, the minimum mass for XTE J1550-564 is defined as
$(9.5 - 1.1)*f_{min} = 8.4\times 125 = 1050$ $M_{\odot}$. See \S 4 for
additional discussion.}  \tablenotetext{e}{Relevant references for the
mass and QPO frequency measurements: (1) Orosz et al. 2002; (2)
Vignarca et al. 2003; (3) ST09; (4) Greiner, Cuby \& McCaughrean 2001;
(5) McClintock et al. 2009; (6) Fillipenko \& Chornock 2001; (7)
Zurita et al. 2002; (8) Munoz-Darias et al. 2008; (9) Hynes et
al. 2004.  measurements.}  \tablenotetext{f}{There is no dynamical
mass constraint for this source.}
\end{deluxetable}

\end{document}